\documentclass[manuscript]{aastex}
\begin{document}
\tighten
\title{ Proper Motions of the HH~47 Jet Observed with the Hubble Space Telescope \altaffilmark{1}}

\author{
	Patrick Hartigan \altaffilmark{2}, 
	Steve Heathcote \altaffilmark{3}, 
	Jon A. Morse \altaffilmark{4},
        Bo Reipurth \altaffilmark{5}, \\
	and John Bally \altaffilmark{6}
	}

\vspace{1.0cm}

\altaffiltext{1}{Based on observations made with the NASA/ESA
{\it Hubble Space Telescope}, obtained at the Space Telescope Science
Institute, which is operated by the Association of Universities for
Research in Astronomy, Inc., under NASA contract NAS5-26555.}

\altaffiltext{2}{Dept. of Physics and Astronomy, Rice University, 
6100 S. Main, Houston, TX 77251-1892}

\altaffiltext{3}{Cerro Tololo Interamerican Observatory, NOAO, Casilla 603, 
La Serena, Chile} 

\altaffiltext{4}{Dept. of Physics and Astronomy, Arizona State University, Tempe AZ } 

\altaffiltext{5}{Institute for Astronomy, University of Hawaii, 640 N. Aohoku Place, Hilo HI 96720} 

\altaffiltext{6}{Center for Astrophysics and Space Astronomy, 389 UCB,
University of Colorado, Boulder, CO 80309, USA}

\begin{abstract}

We present a proper motion study of the shock waves
within the classic stellar jet HH~47 based on Hubble Space Telescope
H$\alpha$ and [S~II] images of the region taken over two epochs.
Individual knots within the jet and in the bow shock/Mach disk working surface of HH~47A
move significantly in the five years that separate the images, and the 
excellent spatial resolution of HST makes it possible 
to measure the proper motions with enough precision to easily observe 
differential motions throughout the flow. The bright portion of the jet emerges at
37.5 $\pm$ 2.5 degrees from the plane of the sky with an average velocity of 300 km$\,$s$^{-1}$.
Dynamical ages of the shock waves in the jet range from a few decades for knots
recently ejected by the source to $\sim$ 1300 years for the faint extended bow
shock HH~47D. The jet curves, but motions of knots in the jet are directed
radially away from the exciting source, and velocity variability in the
flow drives the shock waves that heat the jet internally.
The jet orientation angle varies with time by about 15 degrees,
and currently points to the northwestern
portion of a cavity outlined by a reflection nebula, where a quasi-stationary
shock deflects the jet. The major working surface HH~47A is more complex than a
simple bow shock/Mach disk, and contains numerous clumps that move relative
to one another with velocities of $\sim$ $\pm$ 40 km$\,$s$^{-1}$. Small clumps
or instabilities affect the Mach disk, and dense clumps may move all the
way through the working surface to cause the bumpy morphology seen at the bow shock.
A localized area between the bow shock and Mach disk 
varies significantly between the two sets of images.


\end{abstract}

\section{ INTRODUCTION  }

Since their discovery in the mid-1980's, collimated jets from young stars
have played an increasingly important role in clarifying how
stars form. Jets provide a means for protoplanetary disks to shed angular momentum
as disk material accretes onto the star, and recent observations of rotation in jets
provide a direct observational link to the accretion disk that ultimately powers 
these flows \citep{coffey04}.  Most stars are members of binaries,
and jets provide crucial information about
disks in such systems through measurements of precession \citep{gomez97}, and
by tracing separate flows from both components of close
binaries \citep{reipurth99,hk03}.  Jets evacuate large cavities that
pierce the surrounding molecular cloud and power large-scale bipolar molecular
outflows which help to support molecular clouds against gravitational collapse \citep{rb01}.
Because outflows are driven by accretion, knots in jets give a fossil record of the accretion
history of young stars.  Collimated outflows also accompany accretion disks in compact galactic sources \citep{mirabel99},
and in extragalactic nuclei \citep{zensus97}; jets from young stars are a particularly good place to
study the connection between accretion and outflow because shock waves in the flow cool
radiatively, so images and spectra reveal where the flows are heated and
trace how they propagate.  

The excellent spatial resolution provided by the Hubble Space Telescope
has greatly advanced our understanding of jet dynamics.
Cooling zones behind shock waves
in stellar jets are typically near the limit of what can be resolved from
the ground, but these zones are clearly resolvable with {\it HST}. 
Shock waves in nearby jets move $\sim$ 1 arcsecond per decade, but also change their
spatial structure on that timescale. Hence, the subarcsecond spatial resolution 
provided by {\it HST} makes it possible to detect real differential motions between
jet knots, and thereby observe the complex dynamics of working
surfaces, merging bow shocks, and fluid instabilities in real time. 
It is possible to observe the true density structure and calculate mass loss rates 
directly when jets are irradiated by ultraviolet light \citep{yz05}. 

The three bright stellar jets (HH~1, HH~34, and HH~111) we have observed to date in this manner
each reveal something new and unexpected about the dynamics of jets from young stars.
A portion of the large bow shock HH~34 \citep{reipurth02} appears to be in the process of
forming a Rayleigh-Taylor instability, and we observed new shock waves form as dense knots
emerged from the source.  The proper motions in HH~111 \citep{hartigan01} clearly show
the aftermath of merging bow shocks, and faint emission
wings along the shocks in the flow demonstrate that the surrounding
material plays no role in collimating the jet once the jet emerges from the source.
Because HH~111 is in the plane of the sky, the proper motions are the true space velocities,
so here we can observe the actual velocity variations in the flow and thereby test
models of radiative shocks in jets.
A remarkable area of strong shear exists along the edges of the bow shock HH~1,
while the flow is much more complex and less ordered on the opposite side of the
flow in HH~2 \citep{bally02}.

Discovered by \cite{schwartz77}, the famous HH~46/47 bipolar outflow
is in many ways an ideal object to study jet dynamics. The jet is bright,
and shows a classic structure of a collimated flow with several large bow shocks.
The brightest bow shock, HH~47A, is one of the best examples of a working surface,
and H$\alpha$ images of the system taken with {\it HST} \citep[][hereafter H96]{heathcote96} 
reveal a distinct bow shock that accelerates downstream material and a Mach disk that
decelerates incident jet gas.
The star formation in the region is relatively simple -- the
jet emanates from a resolved ($\sim0\farcs26$) low mass binary system surrounded
by a circumstellar disk \citep{reipurth00} that has formed in the isolated
Bok globule ESO 210-6A at the edge of the Gum Nebula. This binary is the only known
young stellar system in the globule.  The globule is clearly compressed to the north
and west toward the source of ionizing radiation (Figure~1 of H96).
The nascent stellar system and circumstellar disk are thus transitioning from
being embedded in the dense globule to being exposed to the UV 
radiation from the nearby massive stars $\zeta$ Pup and $\gamma^2$ Vel
through photoevaporation of the surrounding envelope \citep{r83}.

The HH~46/47 outflow has been studied extensively across the
electromagnetic spectrum at UV, optical, and infrared wavelengths
\citep[H96;][]{hartigan99,rh91,chernin91,eis94,nc04}.
The optically bright HH~46/47 jet is part of the
northeast flow, and terminates in the HH~47A bow shock
$\sim1.3^\prime$ from the IR driving source. HH~47A moves outward
in the wake of the larger, fainter bow shock HH~47D. 
A faint counterjet in the redshifted southwest lobe \citep{em94}
extends toward the bow shock HH~47C \citep{dopita82}. 
Assuming a distance of 450 pc \citep{egg80,claria82}, the
visible outflow from HH~47D to HH~47C extends $\sim$0.57 pc. The
outflow velocities of $\sim$300 km$\,$s$^{-1}$ derived from
proper-motion and radial-velocity studies \citep{hrm90,rh91,em94,micono98,sg03}
imply a kinematical age for the main optical/near-IR outflow
of $\sim10^3$ years, with large mass ejections occuring every 400 years
or so \citep{morse94}, perhaps driven by FU Orionis-type eruptions \citep{dopita78}.
However, as occurs with other HH flows \citep[e.g. HH 34][]{bd94}, 
deep wide field images reveal that HH~46/47 extends at least 2.6~pc from the IRS, and
therefore must be $\gtrsim 10^4 - 10^5$ years old \citep{stanke99}.

In this paper we continue our investigation of stellar jet dynamics by presenting
{\it HST} proper motions of the optically emitting shocks in the HH~46/47 jet. 
The new observations were taken about 5 years after those in H96, long 
enough for knots in the jet to move significantly, but short enough that morphological
changes remain small for most of the knots. In addition to being able to quantify
precession and differential motions in the jet with high precision, the new images
make it possible to study the dynamics of a working surface in exquisite detail
for the first time. 

\section{OBSERVATIONS AND DATA REDUCTION}

We imaged the HH~47 jet through the F656N and the F673N
filters with the WFPC-2 camera on {\it HST} on 26 February 1999, 
a time interval of 4.9 years after the first epoch of images described by H96.
The images include the entire bright jet on one WF chip, with the extended
bow shock HH~47D spread across the PC (Fig.~1).
Three exposures each for H$\alpha$ and [S~II] facilitated cosmic-ray
removal. Total exposure times for the H$\alpha$ and [S~II] images in the 1999 epoch
were 4100 seconds and 4000 seconds, respectively.
We reduced the new images and aligned them with the previous epoch using the
same procedures described in H96. 

A difference image between the two epochs (Fig.~2) shows clearly that the
jet and bow shocks move away from HH~47-IRS.
To measure proper motions we used a method described by \cite{currie96},
explained in more detail in \cite{hartigan01}.
The algorithm shifts two images to minimize the
square of the difference of the counts over a region that defines the object.
We examined each image for features that retained their overall shape  
between the two epochs and applied the procedure 
separately for H$\alpha$ and for [S~II], because HH objects
often have different morphologies when imaged in these two filters.
Figs.~3 and~4 show objects in the jet, while Figs.~5 and~6 cover the HH~47A
bow shock. Variable features within HH~47A are emphasized in Fig.~7.
Table~1 compiles proper motions of all the objects identified in the figures.
Coordinates in the table were obtained using the stsdas task xy2rd.

Errors in the proper motions come from imperfect alignment of the images between
epochs, uncertainties in the centroiding routine that determines the spatial shifts,
and errors in the distance to HH~47.
In the HH~47 images there are many field stars available to align the
epochs, so uncertainties in the registration are small.
Using the proper motion code on stars after the images were
aligned shows the rms error in the image registration is about 0.1 pixel.
Formal statistical errors in centroiding features are typically 0.05 pixel.
However, an additional systematic error of $\sim$ 0.1 pixel arises from using different
buffer sizes to search for the centroid (e.g. H96). The distance to HH~47 derives from
estimates of the distance to NGC~2547 \citep[450 pc][]{claria82} and to a group of coeval stars
within about 5 degrees of the Vela pulsar \citep[425 pc][]{egg80}. The projected distance
of HH~47 from NGC~2547 is about 3 degrees, or 25~pc at a distance of 450~pc. We adopt
25~pc as the uncertainty in the distance, which introduces a 6\%\ error in the observed
tangential velocities.

Altogether, internal errors in the proper motions over the 4.9 year interval between epochs are
$\sim$ 5 km$\,$s$^{-1}$, for all objects except the large, 
diffuse bow shock HH~47D, where the uncertainty is 40 km$\,$s$^{-1}$. The 6\%\ uncertainty in the
distance is a multiplicative constant that affects all the measurements in the same way.
Proper motions in HH~47D derive from the sum of the H$\alpha$ and [S~II] images to improve
the signal-to-noise.

\section{PROPER MOTIONS, VARIABILITY, CLUMPY FLOW, AND THE CAVITY OF THE HH 47 JET}

When combined with existing radial velocity information, our new proper motion measurements
make it possible to calculate the orientation of the flow, its true space velocity, and to
assess the degree to which clumpy flows and variations in the intrinsic velocity and in the
ejection angles affect the jet dynamics. 
Initial results from the analysis of these data were reported by \cite{hartigan03}.
Radial velocities along the jet observed by Meaburn \& Dyson (1987, their Fig.~2)
increase from HH46~IRS out to the HH~47A bow shock, and we observe the same behavior in the
proper motions shown in Fig.~8.  This correlation between radial velocities and proper motions
implies that the jet varies intrinsically in velocity.  The alternative explanation, that
changes in the ejection angles cause the radial velocities and proper motions to change, is
ruled out because in this model proper motions increase as radial velocities 
decrease and vice-versa for a constant velocity flow.

The relatively poor spatial resolution of the ground-based radial velocity measurements makes
it difficult to compare these directly with proper motions of specific emission features observed by {\it HST}.
However, it is easy to identify the maximum radial velocity of the jet with respect to the cloud
as being 200 $\pm$ 10 km$\,$s$^{-1}$ before the jet encounters HH~47A \citep{md87,hrm90}.
We measure the proper motion of knot Ah1 to be 300 $\pm$ 18 km$\,$s$^{-1}$ (where the error is
dominated by uncertainties in the distance), so this feature
moves at an angle of 34 $\pm$ 2 degrees from the plane of the sky, with a velocity
of 360 $\pm$ 28 km$\,$s$^{-1}$.

Because the jet is not straight, a long slit may miss some of the line emission in specific features
and thereby give misleading estimates for the orientation angles.
To obtain radial velocities for the entire emitting region one can use
a slit map or observe with a Fabry-Perot, as was done by
\cite{morse94}.  Using the datacube from that paper, the average radial velocity of the jet
with respect to the cloud is 180 $\pm$ 10 km$\,$s$^{-1}$. The average proper motion of
the jet in Fig.~8 is 235 km$\,$s$^{-1}$, with an intrinsic scatter of 10 km$\,$s$^{-1}$
and a systematic error of 14 km$\,$s$^{-1}$ from the distance.
Hence, the average orientation angle of the jet with respect to the plane of
the sky is 37.5 $\pm$ 2.5 degrees, in good agreement with the above measurement of
Ah1 and with previous estimates from ground-based proper motions \citep{em94}.

As the last column of Table~1 shows, the outer bow shock HH~47D is
about 1300 years old, and HH~47A was ejected a little less than 800 years ago, assuming
no significant deceleration of these objects has occurred. The average proper motion 
is $\sim$ 235 km$\,$s$^{-1}$ in the jet, $\sim$ 200 km$\,$s$^{-1}$ in HH~47A and $\sim$
180 km$\,$s$^{-1}$ in HH~47D, so if this decline is caused by deceleration rather
than source variability, then the amount of deceleration over the length of the jet
is rather small compared with the velocity of the flow.  The youngest knots in the jet are
only a few decades old, and these knots have significantly higher proper motions than those
ejected 150 $-$ 200 years ago.

The HH~47 flow is an excellent place to observe how a jet with a variable ejection angle 
clears out a larger cavity. The inner 20 arcseconds of the HH~47 jet curves to the left
in Figs.~1 and~2, but the proper motions are not along the jet. Instead, the knots in the jet
move away from the source radially, as expected for a jet with a variable ejection angle.
The current direction of the jet, defined by the motion of knots Jh1, Jh2, and Js2,
is more northerly than its average over the past millennium. The reflection nebula
along the cavity is also brighter on the northern side, as if the space between the IRS
and the northern cavity wall has been mostly cleared of dust, and Fig.~2 shows that this
wall brightened between 1994 and 1999.
There is direct evidence that the jet currently collides with stationary material on this side of
the cavity. A morphologically variable, but stationary, linear H$\alpha$ feature noted
by H96 has the spatial position and orientation expected if the jet were
impacting the northern edges of the cavity obliquely. It is now possible to study
deflection shocks in laboratory experiments that propagate jets into crosswinds \citep{lebedev04}.
The HH~46/47 deflection shock is present only in H$\alpha$, and is marked in Fig.~3.

Shocks in stellar jets are caused by faster material overtaking slower material
\citep[e.g.][]{hartigan01}. Typical
velocity variations along the HH~47 jet between adjacent knots
are $\sim$ 20 km$\,$s$^{-1}$, with a few
cases up to 40 km$\,$s$^{-1}$, in agreement with shock models of the emission line
ratios \citep{hartigan94}. Two places in the flow will experience higher shock
velocities in the future -- proper motions in knot Ah1 exceed those of HH~47A 
by about 100 km$\,$s$^{-1}$, so the true impact velocity should be $\sim$
125 km$\,$s$^{-1}$ when this knot enters the working surface in about the year
2040. This impact velocity should be large enough to generate [O~III] line emission.
The other area where proper motions change markedly is close to the source, where
knots Jh1 and Jh2 have proper motions $\sim$ 100 km$\,$s$^{-1}$ higher than Jh3
and other knots ahead in the flow. The first dynamical consequence of these large 
proper motions should be in about 30 years, when these knots encounter the current
position of the deflection shock (Fig.~3).

HH~47A appears as a bow shock with a well-defined Mach disk (Ah4 in Fig.~5),
and models of HH~47A have attempted to predict the emission line ratios and line
widths using this scenario.  Based on the lack of [O~III] $\lambda$5007, the shock velocities in
HH~47A are $\lesssim$ 90 km$\,$s$^{-1}$ \citep{morse94}, and the bow shock model that fits
the line widths and the entire optical/UV spectrum the best is one where the bow
shock has a velocity of
60 km$\,$s$^{-1}$ and the Mach disk 40 km$\,$s$^{-1}$ \citep{hartigan99}.
While these shock velocities are generally consistent with the new proper motion data, it
is clear that the dynamics within HH~47A is more complex than a simple bow shock/Mach disk
structure. Figs.~5, 6, and~8 show that individual clumps within HH~47A have different proper
motions, and spread $\pm$ 40 km$\,$s$^{-1}$ or so relative to the mean motion of HH~47A.
For example, knot Ah8 clearly moves faster than the adjacent knot Ah10 (Fig.~7), and Table~1
confirms a relative velocity of $\sim$ 40 km$\,$s$^{-1}$ for these two knots. 

Arrows in Fig.~7 mark two areas on the Mach disk that have changed over the five years
between the epochs. The changes could result from some sort of fluid instability, or
could simply arise when a small dense blob like Ah8 plunges through the Mach disk. The area between
the bow shock and the Mach disk is particularly variable in Fig.~7. In general,
H$\alpha$, where the flux is determined largely by collisional excitation at the shock
and so is proportional to the mass flux across the shock at any given moment,
traces morphological changes more quickly than does [S~II], which radiates in a spatially
extended zone that averages changes out over a cooling time of tens of years.

\section{CONCLUSIONS}

We have measured precise proper motions from two epochs of {\it HST} images of the prototypical stellar
jet HH~47. Uncertainties of the relative proper motions between knots
are typically $\pm$ 5 km$\,$s$^{-1}$, considerably 
lower than variations of the internal motions of the knots in the jet, so we can assess how the
fluid dynamics operates in this system accurately for the first time.

Taking the source to be at a distance of 450~pc, the orientation angle of the flow is
37.5 $\pm$ 2.5 degrees with respect to the plane of the sky. The bright portion of the jet is
about 800 years old, and has an average space velocity of 300 km$\,$s$^{-1}$. 
Although the jet curves, the motion is primarily directed radially away from the source.
The ejection angle of the jet varies with time by $\pm$ 15 degrees or so, and 
currently impacts the northern side of the reflection nebula, where an oblique shock
redirects the supersonic flow.

Proper motions in the jet increase with distance along the bright section of the jet
in front of the bow shock HH~47A, similar to the behavior of radial velocities. This
correlation implies that the source varies intrinsically in velocity, and it is these
velocity variations that produce shocks that heat the jet internally. The observed
variations in proper motion along the jet are consistent with shock models that were
designed to explain the observed emission line ratios in the jet.

The bright bow shock HH~47A has a clear Mach disk and bow shock, but the object is
more complex than a dual shock structure, and has many clumps whose differential motions
dominate much of the dynamics in this object. The region between the bow shock and Mach
disk has a highly variable morphology, and the Mach disk also changed structure significantly
over the five year interval between images.  The individual knots within HH~47A are
easiest to explain as small bullets that pass from the jet
through the Mach disk and working surface to emerge as bumps in the bow shock. 

Understanding HH flow dynamics will require 
systematic work that spans decades, but the rewards are great because these objects
make it possible to test our basic understanding of the time dependences of shocked
supersonic flows, a ubiquitous phenomenon in astrophysics. 
Additional epochs of high spatial resolution imaging of HH~47 will be able to follow the dynamics
of this fascinating object in real time, and thereby provide ideal constraints for
numerical models of working surfaces within collimated supersonic flows.

\acknowledgements
{Support was provided by NASA through {\it HST} program GO-6794
from the Space Telescope Science Institute, which is operated
by the Association of Universities for Research in Astronomy, Inc.,
under NASA contract NAS~5-26555.}

{}

\newpage

\begin{center}
\begin{planotable}{cccrrrrrr}
\tablecolumns{9}
\tablewidth{470pt}
\tablecaption{Proper Motions in HH 47}
\tablehead{
\colhead{Object\tablenotemark{a}} &\colhead{$\alpha$ (2000)\tablenotemark{b}} &\colhead{$\delta$ (2000)\tablenotemark{b}}&
\colhead{$\Delta$ X\tablenotemark{c}} &\colhead{$\Delta$ Y\tablenotemark{c}}
&\colhead{V$_{X\perp}$\tablenotemark{d}}&\colhead{V$_{Y\perp}$\tablenotemark{d}}&
\colhead{V$_{\perp}$\tablenotemark{d}}&
\colhead{Age\tablenotemark{e}}
}
\tablenum{1}
\startdata
Jh1     &8:25:44.009&$-$51:00:33.49&117 &36   &249 &76   &261& 19\\
Jh2     &8:25:44.154&$-$51:00:32.27&124 &32   &264 &68   &273& 32\\
Jh3     &8:25:44.826&$-$51:00:25.10& 80 &22   &170 &46   &176&166\\
Jh4     &8:25:45.556&$-$51:00:22.40& 89 &$-$1 &191 &$-$3 &191&229\\
Jh5     &8:25:45.816&$-$51:00:23.60& 95 &6    &203 &13   &204&228\\
Jh6     &8:25:45.863&$-$51:00:21.42&100 &14   &214 &30   &217&231\\
Jh7     &8:25:46.005&$-$51:00:19.93&102 &8    &219 &17   &219&247\\
Jh8     &8:25:46.224&$-$51:00:17.81&103 &1    &220 &2    &220&275\\
Jh9     &8:25:46.629&$-$51:00:14.06& 96 &6    &205 &12   &205&350\\
Jh10    &8:25:47.415&$-$51:00:13.17&111 &15   &237 &31   &239&359\\
Jh11    &8:25:47.215&$-$51:00:08.11& 85 &35   &181 &75   &196&456\\
Jh12    &8:25:48.423&$-$50:59:59.35&129 &10   &275 &20   &276&436\\
Ah1     &8:25:49.783&$-$50:59:52.42&139 &16   &297 &34   &299&505\\
$<$Ah$>$&8:25:50.262&$-$50:59:52.51& 93 &$-$2 &199 &$-$4 &199&797\\
Ah2     &8:25:50.146&$-$50:59:53.65& 90 &$-$12&192 &$-$26&194&\\
Ah3     &8:25:49.978&$-$50:59:50.69&103 &10   &221 &22   &222&\\
Ah4     &8:25:50.156&$-$50:59:52.45& 96 &$-$7 &206 &$-$16&206&\\
Ah5     &8:25:50.260&$-$50:59:52.87& 79 &6    &169 &12   &169&\\
Ah6     &8:25:50.466&$-$50:59:53.37& 94 &$-$14&201 &$-$31&203&\\
Ah7     &8:25:50.208&$-$50:59:50.34& 89 &$-$8 &189 &$-$16&190&\\
Ah8     &8:25:50.307&$-$50:59:51.08&115 &$-$4 &244 &$-$8 &245&\\
Ah9     &8:25:50.360&$-$50:59:51.84& 85 &$-$5 &182 &$-$11&183&\\
Ah10    &8:25:50.304&$-$50:59:50.42& 96 &$-$10&204 &$-$21&205&\\
Ah11    &8:25:50.424&$-$50:59:51.51& 91 &$-$8 &195 &$-$17&195&\\
Ah12    &8:25:50.409&$-$50:59:50.77& 98 &$-$9 &210 &$-$18&211&\\
Js1     &8:25:44.017&$-$51:00:33.83&125 &32   &267 &68   &276& 16\\
Js2     &8:25:44.816&$-$51:00:25.12& 88 &20   &187 &43   &192&151\\
Js3     &8:25:45.011&$-$51:00:24.17&114 &18   &243 &38   &246&135\\
Js4     &8:25:45.312&$-$51:00:23.18&115 &6    &245 &13   &245&159\\
Js5     &8:25:45.445&$-$51:00:21.19& 96 &3    &204 &7    &204&215\\
Js6     &8:25:45.532&$-$51:00:22.39& 97 &$-$2 &207 &$-$3 &207&210\\
Js7     &8:25:45.811&$-$51:00:23.53& 94 &5    &200 &10   &200&232\\
Js8     &8:25:45.688&$-$51:00:21.16&102 &$-$7 &219 &$-$15&219&217\\
Js9     &8:25:45.870&$-$51:00:21.43&102 &15   &218 &32   &220&227\\
Js10    &8:25:46.031&$-$51:00:19.25&108 &8    &231 &17   &231&240\\
Js11    &8:25:46.319&$-$51:00:17.95&108 &2    &230 &5    &230&269\\
Js12    &8:25:46.697&$-$51:00:15.18&111 &$-$2 &236 &$-$5 &236&303\\
Js13    &8:25:46.951&$-$51:00:15.06&114 &$-$2 &244 &$-$5 &244&311\\
Js14    &8:25:47.274&$-$51:00:11.89&107 &10   &229 &21   &230&370\\
Js15    &8:25:47.284&$-$51:00:08.15&116 &3    &248 &7    &248&364\\
Js16    &8:25:47.655&$-$51:00:08.53&122 &15   &260 &32   &262&365\\
Js17    &8:25:47.544&$-$51:00:06.02&118 &5    &251 &12   &251&387\\
Js18    &8:25:48.048&$-$51:00:02.95&118 &4    &252 &9    &252&434\\
Js19    &8:25:48.482&$-$51:00:00.36&118 &8    &251 &16   &252&476\\
Js20    &8:25:49.335&$-$50:59:55.84&133 &1    &283 &2    &283&492\\
As1     &8:25:49.783&$-$50:59:52.42&140 &$-$1 &299 &$-$1 &299&505\\
$<$As$>$&8:25:50.254&$-$50:59:52.57& 95 &$-$5 &203 &$-$12&203&779\\
As2     &8:25:50.199&$-$50:59:56.11& 76 &$-$4 &162 &$-$8 &162&\\
As3     &8:25:50.316&$-$50:59:56.31& 92 &5    &196 &11   &197&\\
As4     &8:25:50.284&$-$50:59:54.73& 83 &3    &176 &6    &176&\\
As5     &8:25:50.390&$-$50:59:55.87& 92 &$-$1 &197 &$-$2 &197&\\
As6     &8:25:50.379&$-$50:59:54.87& 90 &$-$7 &192 &$-$15&193&\\
As7     &8:25:50.550&$-$50:59:54.54& 91 &$-$1 &195 &$-$2 &195&\\
As8     &8:25:50.477&$-$50:59:56.62& 83 &$-$7 &178 &$-$15&179&\\
As9     &8:25:50.448&$-$50:59:54.45& 98 &$-$6 &210 &$-$12&210&\\
As10    &8:25:50.475&$-$50:59:53.65& 93 &$-$8 &198 &$-$18&199&\\
As11    &8:25:50.064&$-$50:59:50.49&102 &6    &218 &13   &218&\\
As12    &8:25:50.187&$-$50:59:49.76& 86 &3    &184 &6    &184&\\
As13    &8:25:50.165&$-$50:59:54.43 &94 &$-$4 &201 &$-$9 &201&\\
As14    &8:25:50.118&$-$50:59:53.67&108 &6    &230 &13   &230&\\
As15    &8:25:50.269&$-$50:59:52.92& 85 &8    &182 &17   &183&\\
As16    &8:25:50.380&$-$50:59:51.87& 86 &$-$13&184 &$-$28&187&\\
As17    &8:25:50.417&$-$50:59:51.27& 99 &$-$4 &211 &$-$8 &211&\\
As18    &8:25:50.328&$-$50:59:51.04&111 &$-$6 &236 &$-$12&236&\\
As19    &8:25:50.314&$-$50:59:50.40&101 &$-$11&216 &$-$24&217&\\
As20    &8:25:50.215&$-$50:59:51.59& 67 &10   &143 &21   &144&\\
D       &8:25:53.64 &$-$50:59:32   & 84 &0    &179 &0    &179&1333\\
\enddata
\tablenotetext{a} {Objects are defined by the boxes drawn in the figures. The labels J and A 
mean the object is in the jet or in HH~47A, respectively, and the letters h and s indicate
the H$\alpha$, and [S~II] images, respectively. The objects $<$Ah$>$ and $<$As$>$ are defined by large
boxes that include all the emission that comprises HH~47A in the H$\alpha$, and [S~II] images,
respectively.} 
\tablenotetext{b} {Coordinates are for epoch 1994.24, and equinox 2000.}
\tablenotetext{c} {$\Delta$ X and $\Delta$ Y are the proper motions in milliarcseconds 
per year along, and perpendicular to the axis of the jet, respectively. Positive values of
$\Delta$ Y indicate motions to the right (to the NW) in the figures. The position angle of the
jet on the sky is 54.18 degrees.}
\tablenotetext{d} {Tangential velocity V$_{X\perp}$ along the x$-$direction, V$_{Y\perp}$
along the y$-$direction, and the total proper motion V$_{\perp}$ are in km$\,$s$^{-1}$,
assuming a distance of 450~pc. Uncertainties are $\pm$ 5 km$\,$s$^{-1}$ for all objects
except D, where they are $\pm$ 40 km$\,$s$^{-1}$. An additional systematic
uncertainty of $\lesssim$ 6\%\ may arise from errors in the distance.}
\tablenotetext{e} {Time in years for the object to move from the source to
its present location at its current velocity. The source position, $\alpha$(2000)
= 08:25:43.91, $\delta$(2000) = $-$51:00:35.6, is from an IRAC 2 Spitzer image
(Noriega-Crespo 2005, personal communication).}
\end{planotable}
\end{center}
\vfill\eject


\begin{figure}
\epsscale {0.50}
\plotone{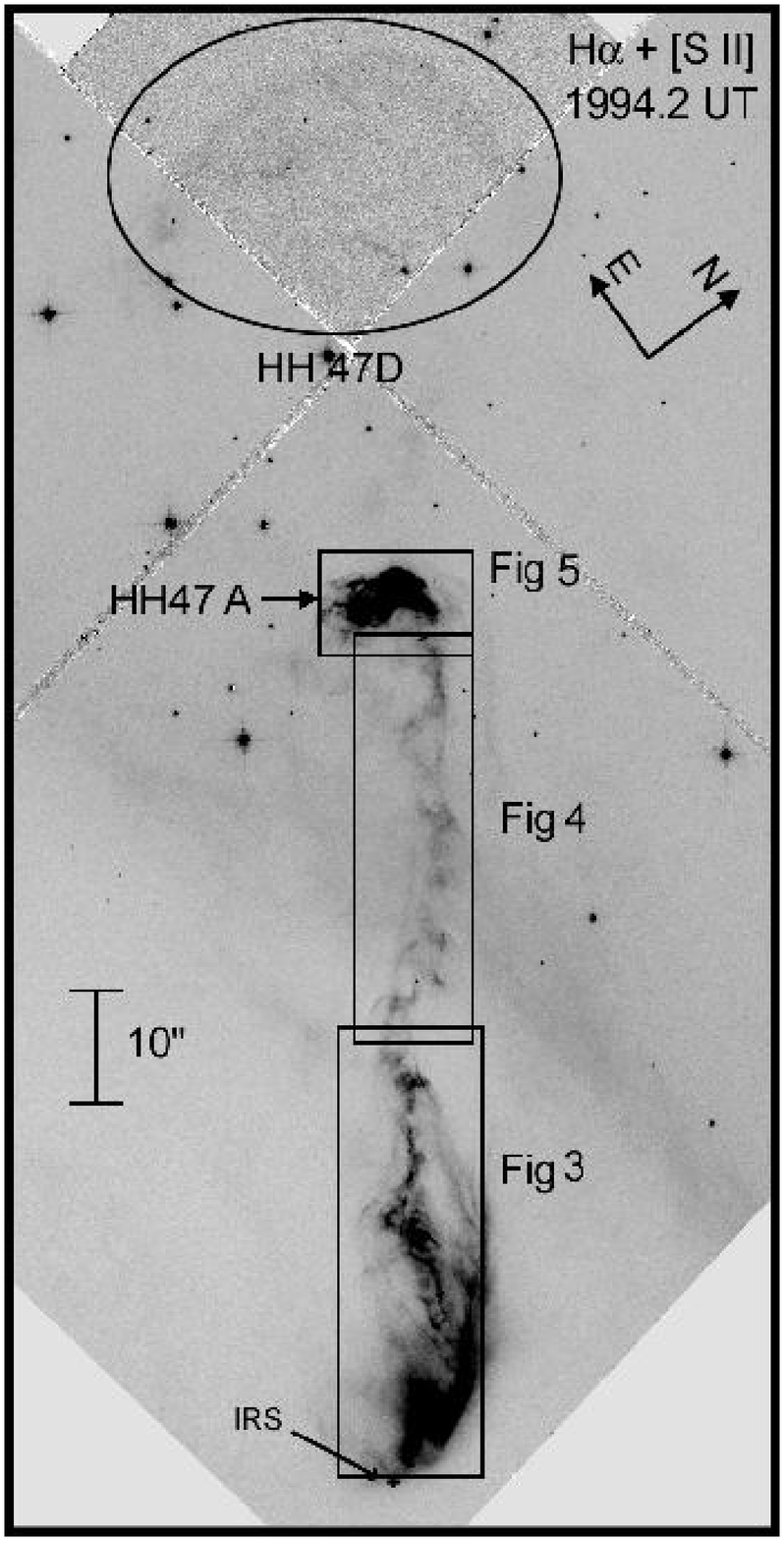}
\caption{An image of the HH 47 jet created by adding together the
[S II] and H$\alpha$ WFPC2 images taken in 1994. The bow shocks HH~47D and HH~47A
are indicated, and boxes show the areas covered by subsequent figures. A
cross marks the position of the exciting source from the
IRAC 2 Spitzer image (2005, Noriega-Crespo personal communication).
The image is rotated to
make the axis of the jet at PA = 54.18$^\circ$ vertical.
\label{fig1}}
\end{figure}
\eject

\begin{figure}
\epsscale {1.00}
\plotone{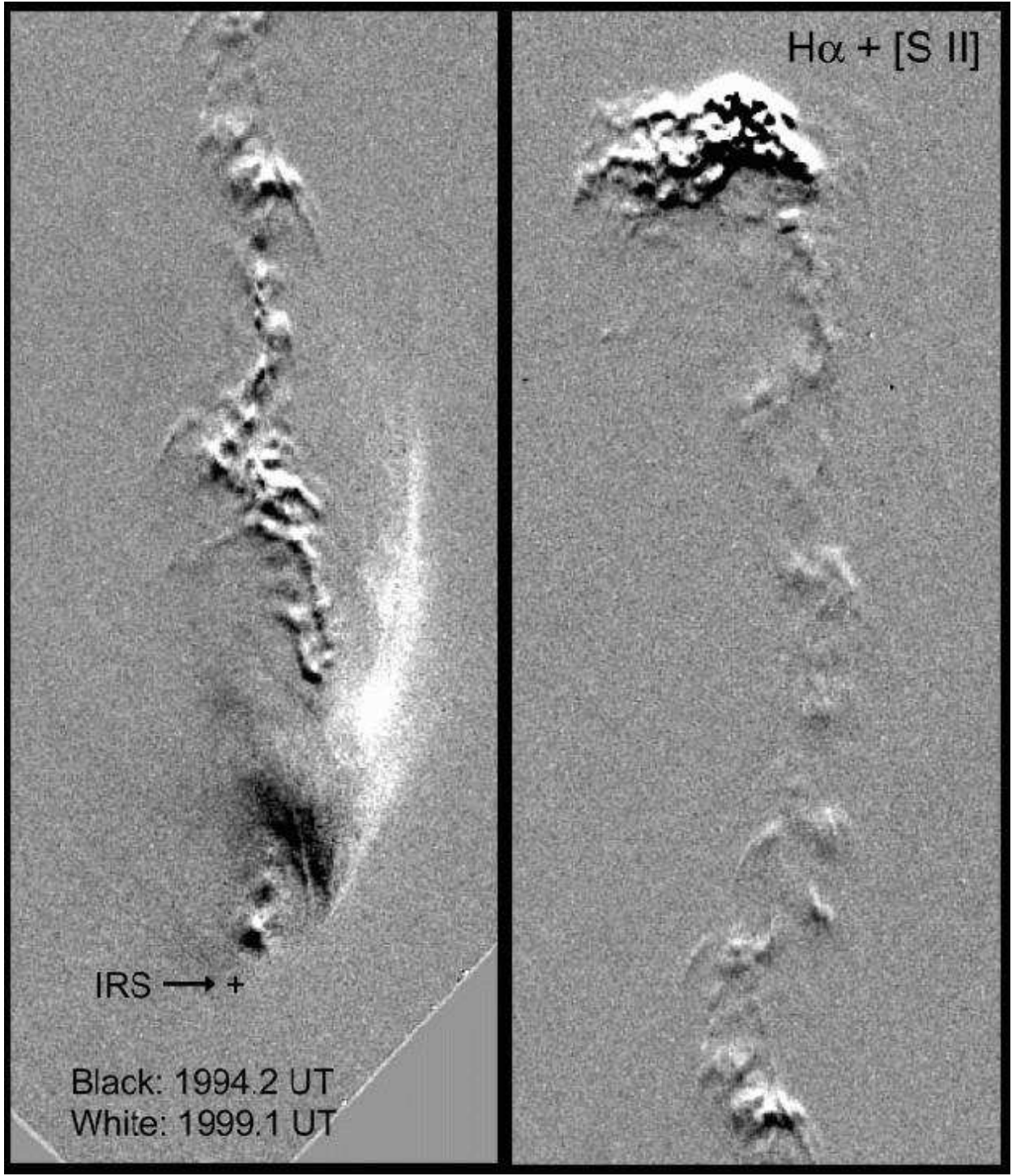}
\caption{Difference between the 1994.2 UT (black) and the 1999.1 UT (white)
epochs of the summed H$\alpha$ + [S~II] images of HH~47. Features in the jet
move radially away from the infrared source (IRS), and so show a leading
white edge and a trailing black edge.  The reflection nebula near
the base of the jet varied between the two epochs, becoming
brighter on the northwest (right) side of the cavity in the second epoch.
\label{fig2}}
\end{figure}

\begin{figure}
\epsscale {1.00}
\plotone{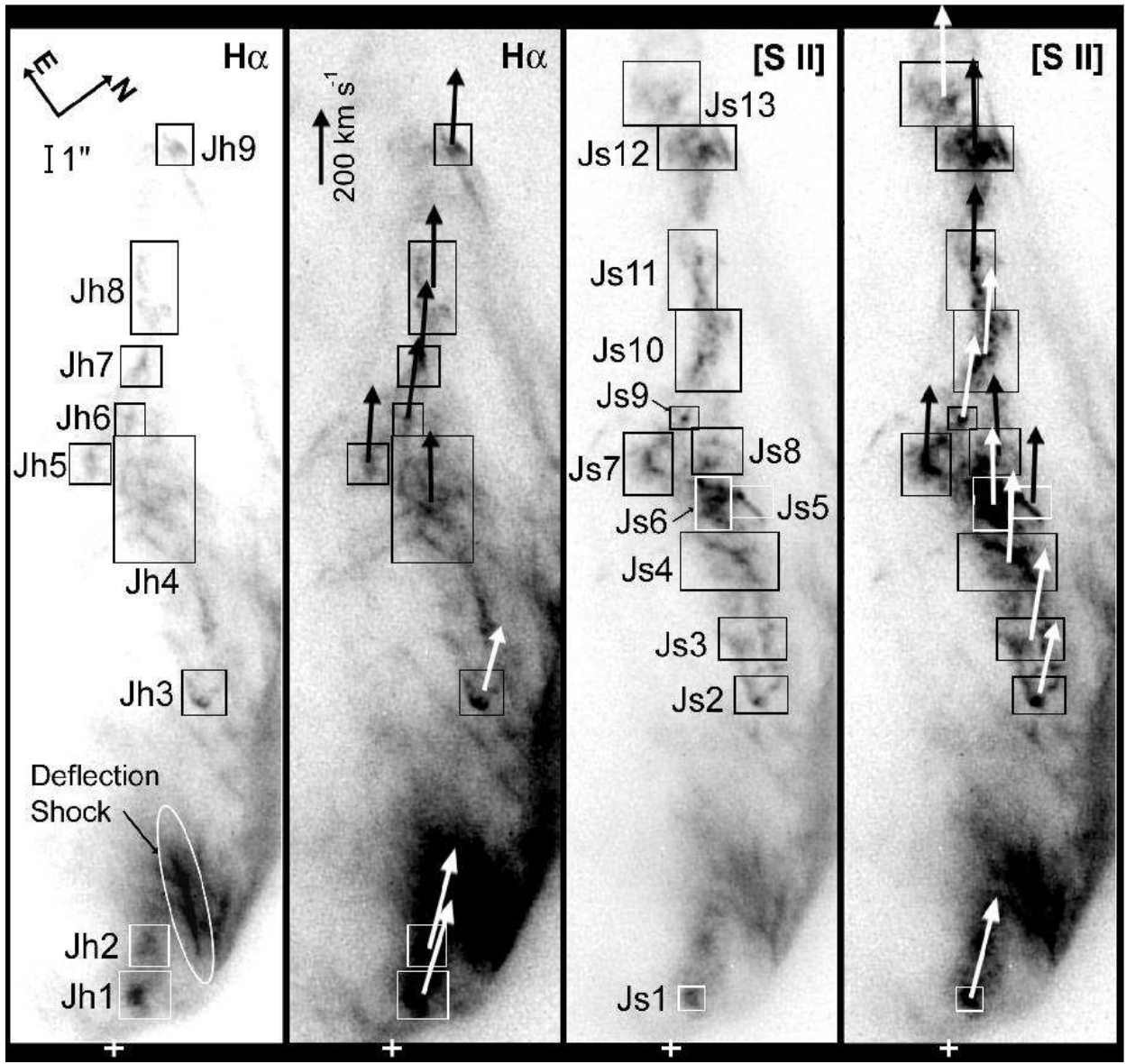}
\caption{Proper motions in the lower section of the HH~47 jet. Boxes
mark features used to measure proper motions in the H$\alpha$ and [S~II]
images, each of which appears at two different greyscales for the 1994 epoch.
Arrows denote the distance each feature would travel in 30 years. The curved
structure on the right side of the image
that extends from Jh1 to Jh9 is a reflection nebula. Crosses denote the
IR source. The figure is oriented as in Fig.~1.
\label{fig3}}
\end{figure}

\begin{figure}
\epsscale {0.70}
\plotone{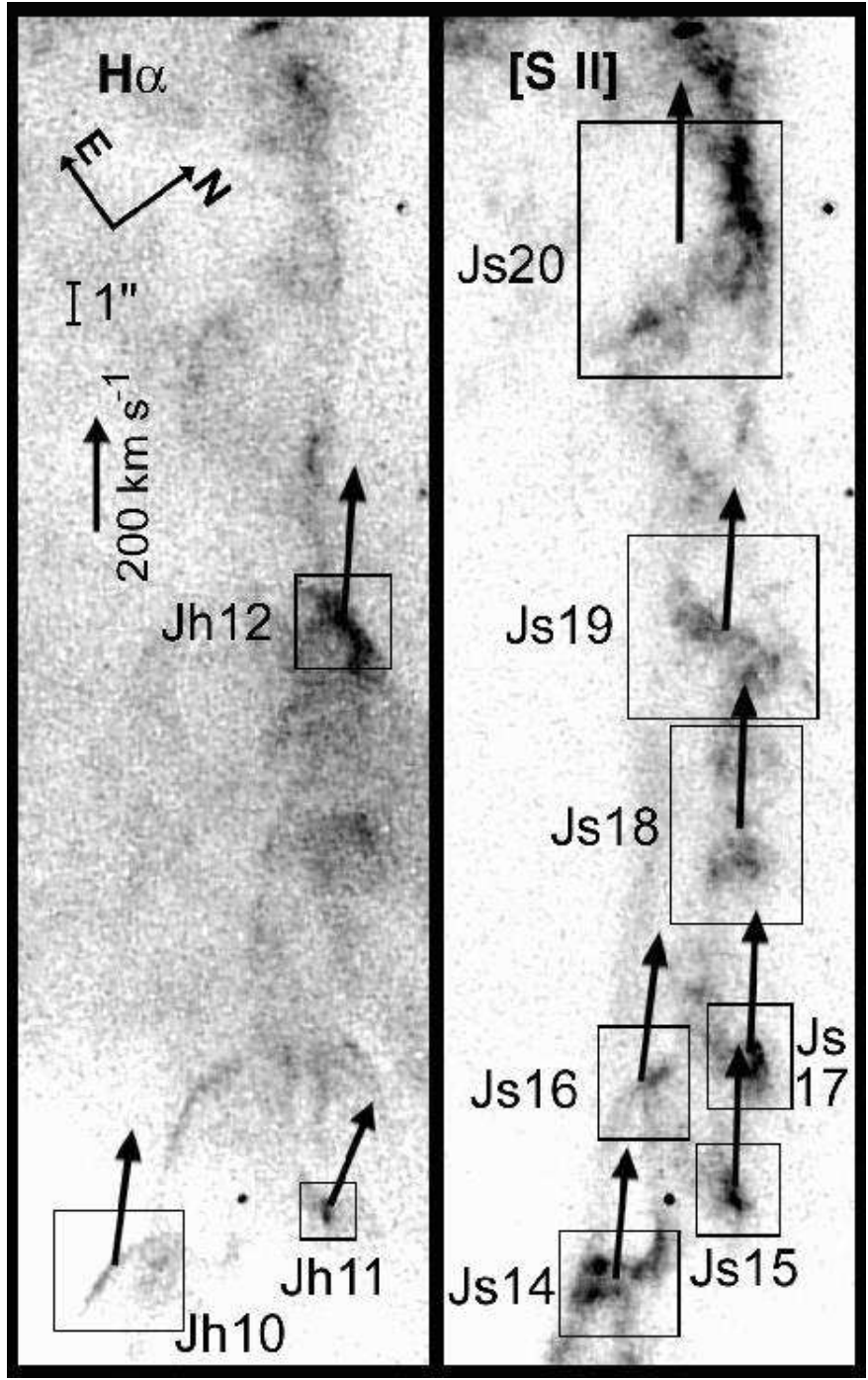}
\caption{Same as Fig.~3 but for the middle portion of the jet.
\label{fig4}}
\end{figure}

\begin{figure}
\epsscale {0.75}
\plotone{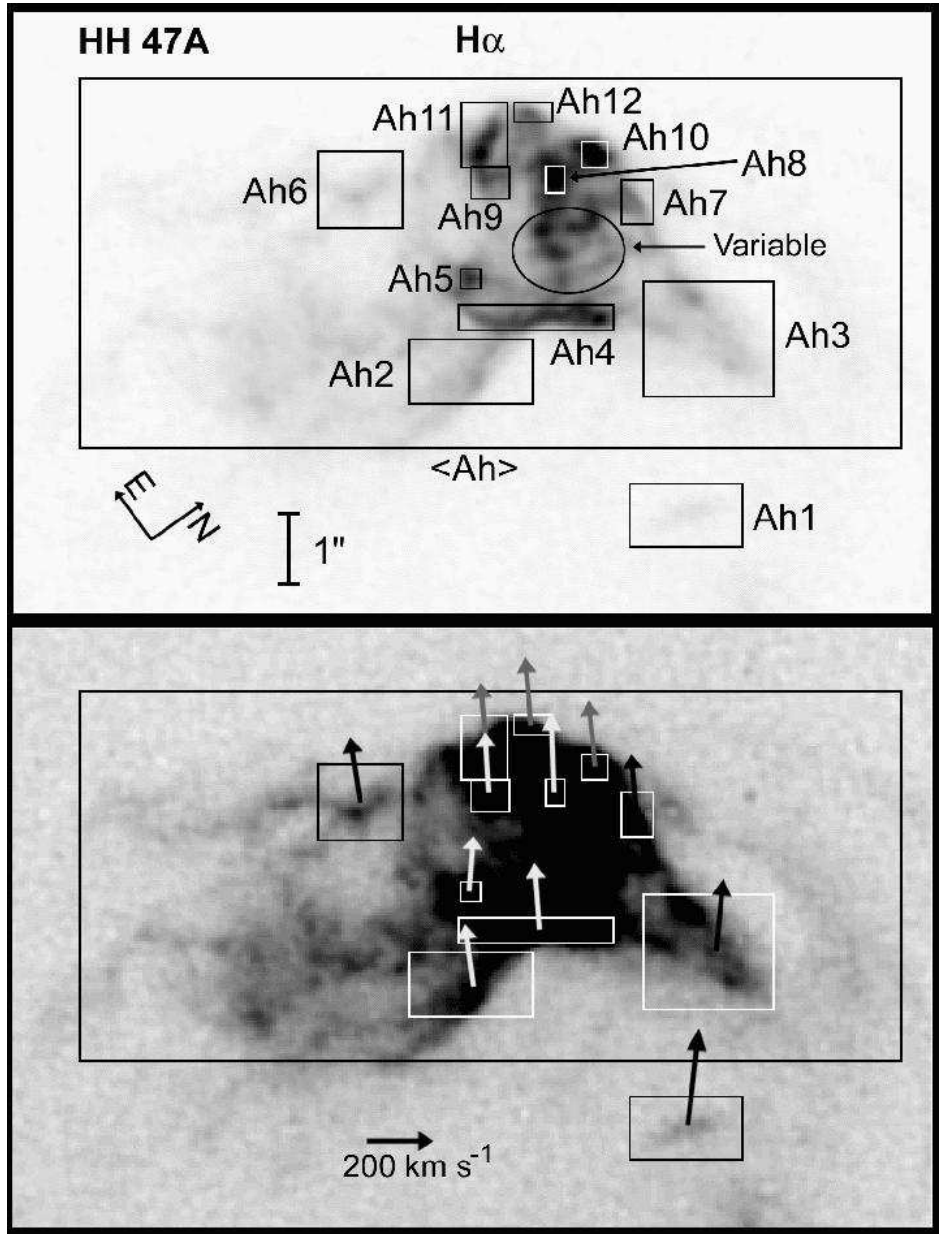}
\caption{Same as Fig.~3 but for two greyscales of the 1994 H$\alpha$ 
image of the HH~47A bow shock. In this
case the arrows show the distance each feature would move in 10 years. The large box
marked $<$Ah$>$ includes emission from the entire structure. There are significant
variations in the velocities between knots (see Table 1). Clumps within the region marked
'Variable' did not retain their shapes well enough to allow proper motion measurements
between the two epochs.
\label{fig5}}
\end{figure}

\begin{figure}
\epsscale {0.75}
\plotone{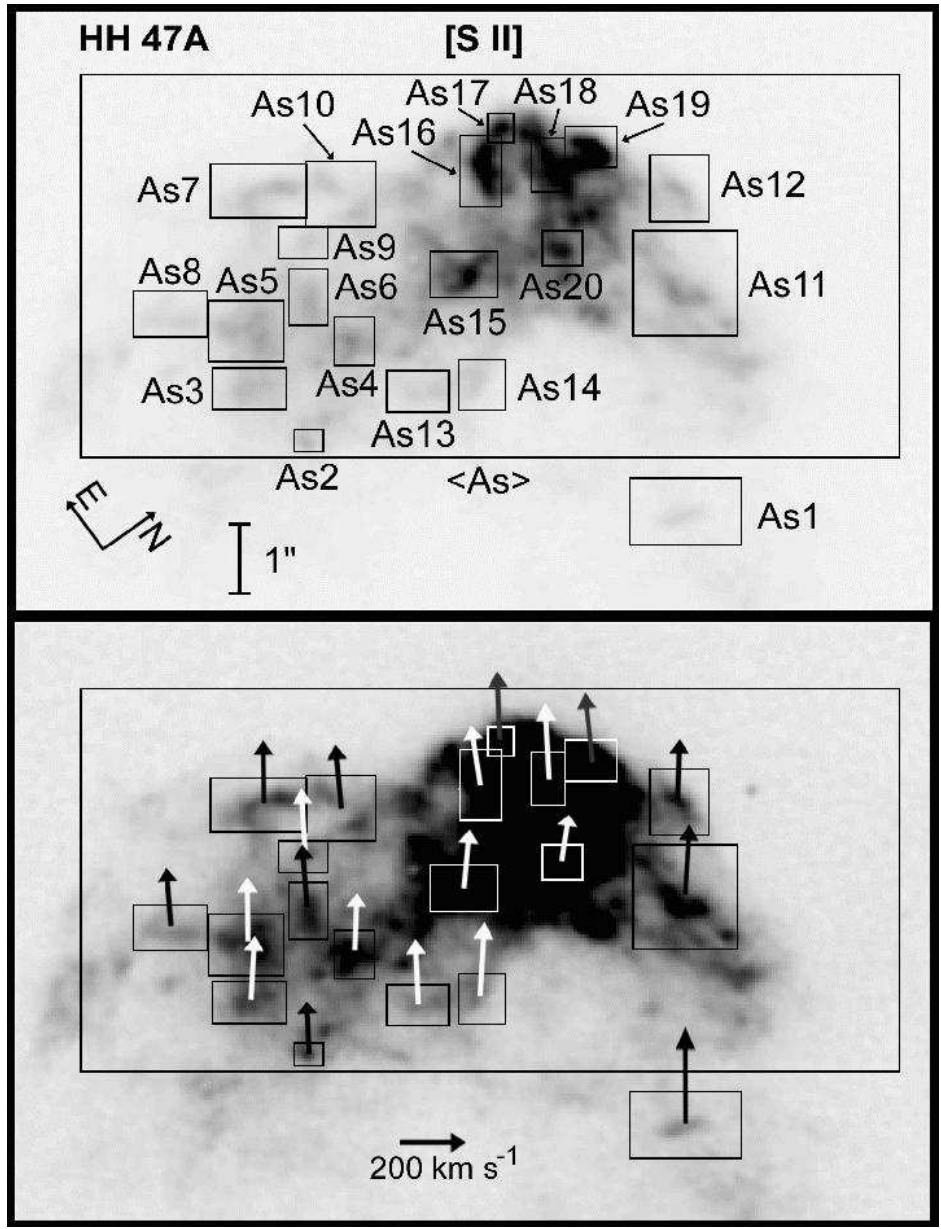}
\caption{Same as Fig.~5 for the HH~47A bow shock in [S~II].
\label{fig6}}
\end{figure}

\begin{figure}
\epsscale {1.00}
\plotone{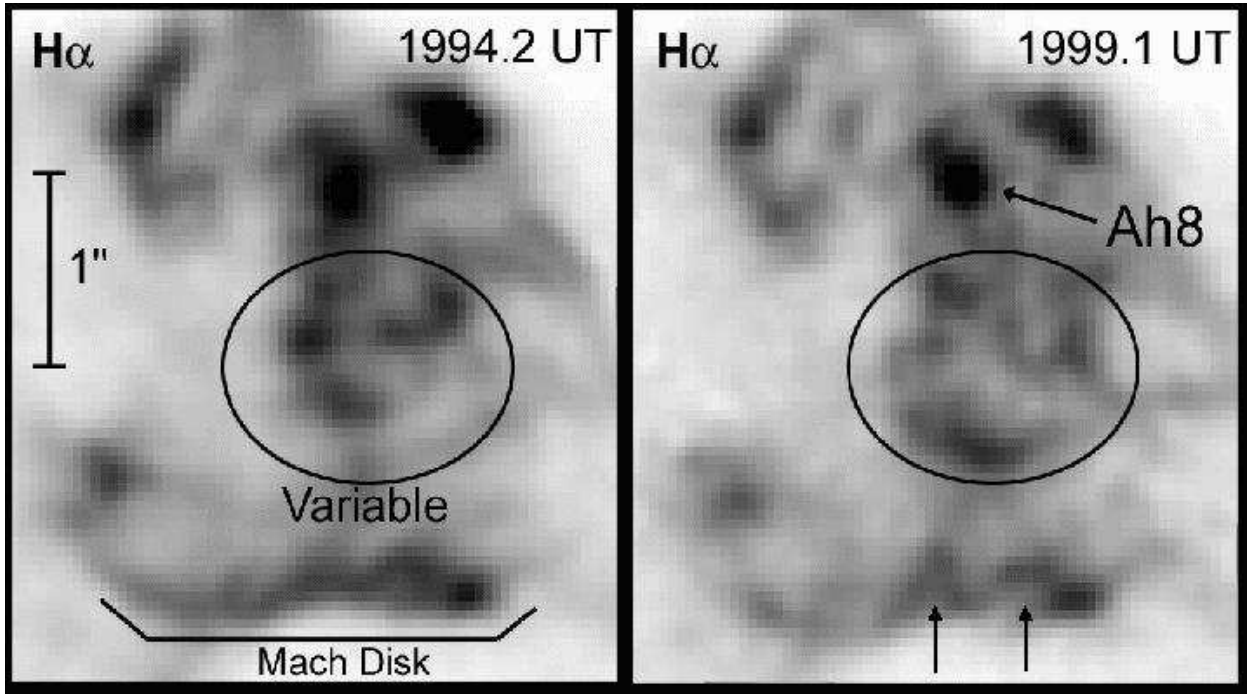}
\caption{An expanded view of the variable region of HH~47A noted in Fig.~5. 
Two arrows at the bottom of the figure
mark locations where the Mach disk changed its structure
significantly. The 1999 epoch was shifted downward by an amount corresponding to
a 200 km$\,$s$^{-1}$ bulk motion over 4.9 years in order to align it with the 1994 epoch. 
The images clearly show the higher velocity of the bright knot Ah8 relative to the 
surrounding emission.
\label{fig7}}
\end{figure}
\normalsize

\begin{figure}
\epsscale {1.00}
\plotone{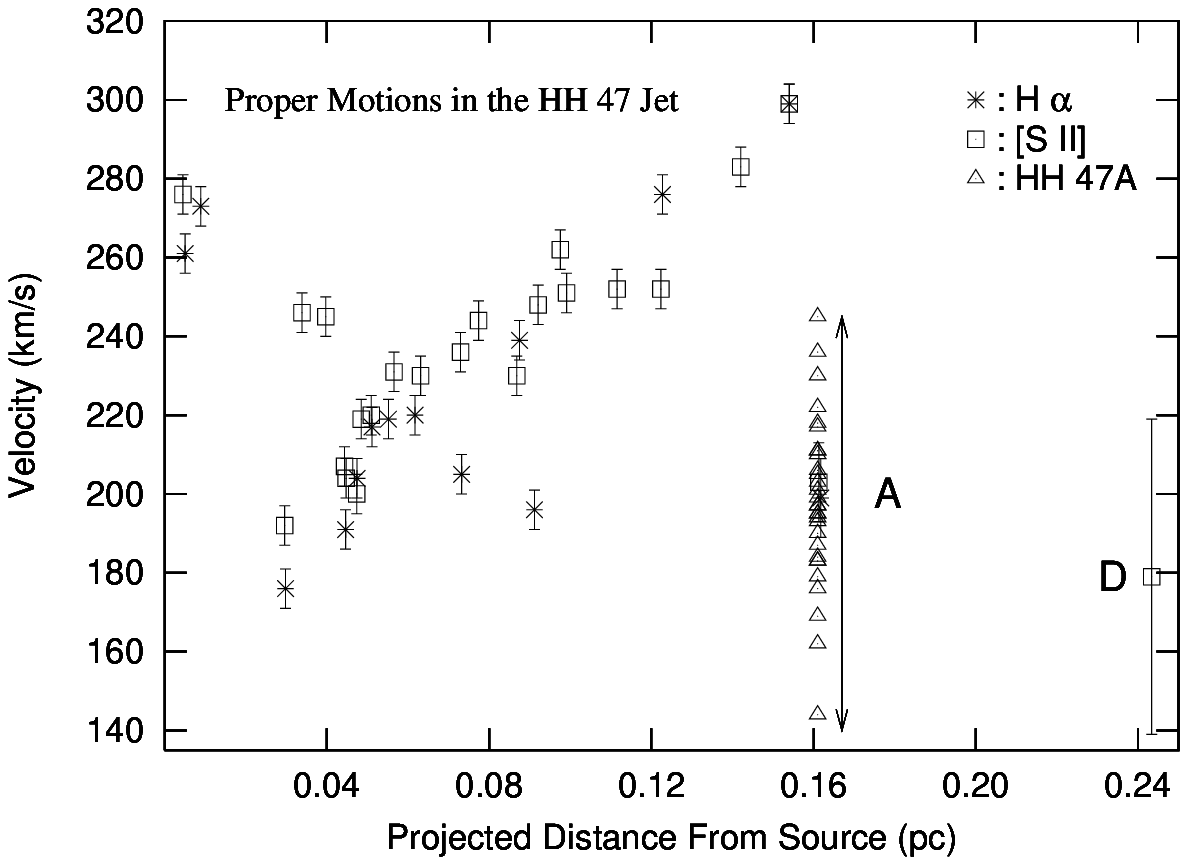}
\caption{}
\label{fig8}
\end{figure}
\normalsize

\end{document}